\documentclass[amsmath,floatfix,twocolumn,superscriptaddress,citeautoscript]{revtex4-1}
\usepackage{subfigure}
\usepackage{siunitx}
\usepackage{amssymb}
\usepackage{amsmath}
\usepackage{graphicx}
\usepackage{array}
\usepackage{dcolumn}
\usepackage{psfrag}
\usepackage{bm}
\usepackage{color}
\usepackage{multirow}

\setcitestyle{super}

\begin{document}

\title{Defect--Limited Efficiency of Pnictogen Chalcohalide Solar Cells}

\author{Cibrán López}
    \affiliation{Departament de Física, Universitat Politècnica de Catalunya, 08034 Barcelona, Spain}
    \affiliation{Barcelona Research Center in Multiscale Science and Engineering, Universitat Politècnica de Catalunya, 08019 Barcelona, Spain}

\author{Seán R. Kavanagh}
    \affiliation{Harvard University Center for the Environment, Cambridge, Massachusetts 02138, United States}

\author{Pol Benítez}
    \affiliation{Departament de Física, Universitat Politècnica de Catalunya, 08034 Barcelona, Spain}
    \affiliation{Barcelona Research Center in Multiscale Science and Engineering, Universitat Politècnica de Catalunya, 08019 Barcelona, Spain}

\author{Edgardo Saucedo}
    \affiliation{Barcelona Research Center in Multiscale Science and Engineering, Universitat Politècnica de Catalunya, 08019 Barcelona, Spain}
    \affiliation{Departament d'Enginyeria Electrònica, Universitat Politècnica de Catalunya, 08034 Barcelona, Spain}

\author{Aron Walsh}
    \affiliation{Thomas Young Centre and Department of Materials, Imperial College London, Exhibition Road, London SW7 2AZ, UK}
    \affiliation{Department of Physics, Ewha Womans University, 52 Ewhayeodae-gil, Seodaemun-gu, Seoul 03760, South Korea}

\author{David O. Scanlon}
    \affiliation{School of Chemistry, University of Birmingham, Birmingham B15 2TT, UK}

\author{Claudio Cazorla}
    \affiliation{Departament de Física, Universitat Politècnica de Catalunya, 08034 Barcelona, Spain}
    \affiliation{Barcelona Research Center in Multiscale Science and Engineering, Universitat Politècnica de Catalunya, 08019 Barcelona, Spain}
    \affiliation{Institució Catalana de Recerca i Estudis Avançats (ICREA), Passeig Lluís Companys 23, 08010 Barcelona, Spain}

\begin{abstract}
Pnictogen chalcohalides (MChX) have recently emerged as promising nontoxic and environmentally friendly photovoltaic absorbers, combining 
strong light absorption coefficients with favorable low-temperature synthesis conditions. Despite these advantages and reported optimized 
morphologies, device efficiencies remain below 10\%, far from their ideal radiative limit. To uncover the origin of these performance losses, 
we present a systematic and fully consistent first-principles investigation of the defect chemistry across the Bi-based chalcohalide family. 
Our results reveal a complex defect landscape dominated by chalcogen vacancies of low formation energy, which act as deep nonradiative 
recombination centers. Despite their moderate charge-carrier capture coefficients, the high equilibrium concentrations of these defects 
reduce the theoretical maximum efficiencies by 6\% in BiSeI and by 10\% in BiSeBr. In contrast, sulfur vacancies in BiSI and BiSBr are 
comparatively benign, presenting smaller capture coefficients due to weaker electron-phonon coupling. Interestingly, despite its huge 
nonradiative charge-carrier recombination rate, BiSeI presents the best conversion efficiency among all four compounds owing to its most 
suitable bandgap for outdoor photovoltaic applications. Our findings identify defect chemistry as a critical bottleneck in MChX solar cells 
and proposes chalcogen-rich synthesis conditions and targeted anion substitutions as effective strategies for mitigation of detrimental vacancies.
\\

\end{abstract}

\maketitle

\section{Introduction}
\label{sec:intro}
The global transition toward sustainable and low-carbon energy systems has placed photovoltaics (PV) at the forefront of research and 
technological innovation \cite{Hermann2006}. As a clean, renewable, and abundant energy source, PV provides a direct route to decarbonizing 
electricity generation. However, photovoltaic technologies must satisfy stringent performance criteria, including high power conversion 
efficiency (PCE), long-term stability, scalability, and low-cost manufacturing, while maintaining environmental sustainability \cite{Peter2011}.

\begin{figure*}[t]
    \centering
        \includegraphics[width=0.8\textwidth]{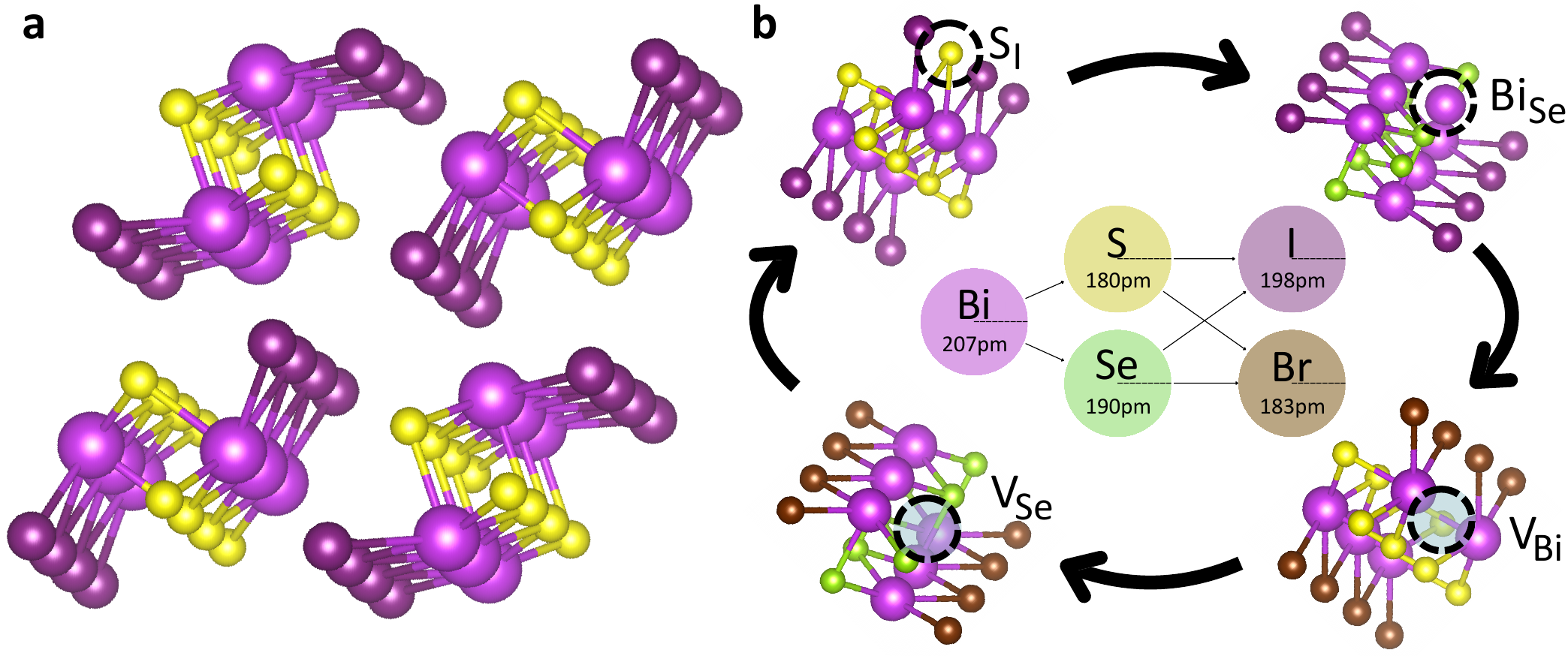}
        \caption{\textbf{MChX photovoltaic absorbers.} \textbf{a.}~MChX orthorhombic crystal structure (space group \textit{Pnma}) characterized 
	by columnar motifs held together by weak dispersion forces. Pnictogen, chalcogen and halide atoms are represented by violet, yellow and 
	dark pink spheres, respectively. \textbf{b.}~A systematic study of MChX point defects disentangles general trends. Four different Bi-based 
	chalcohalides and all possible intrinsic point defects have been systematically analyzed in this study. The van der Waals radius of each 
	species is marked for reference.}
        \label{fig1}
\end{figure*}

Pnictogen chalcohalides (MChX, Fig.~\ref{fig1}a) have recently emerged as a promising class of photovoltaic absorbers owing to their nontoxicity, 
bandgaps in the range of 1.2--2.1~eV, light absorption coefficients spanning from 25 to 66~$\mu$m\textsuperscript{-1} 
\cite{Lopez2024,Ganose2016,Nielsen2025}, exceptional thermodynamic stability \cite{Nie2019,Li2024}, and low synthesis temperatures (below 
$300^\circ$C) \cite{Cano2023,Li2024}. MChX materials are often described as ``perovskite-inspired'' semiconductors \cite{Cano2025}, since their 
electronic structures resemble those of lead-halide perovskites, despite crystallizing in different crystal lattices. In particular, such materials 
are believed to present defect tolerance, stemming from antibonding and bonding character of their valence and conduction bands, respectively. 
Actually, the closely related family of binary chalcogenides (e.g., Sb\textsubscript{2}S\textsubscript{3}, Sb\textsubscript{2}Se\textsubscript{3}, 
and Bi\textsubscript{2}S\textsubscript{3}) has been shown to exhibit genuine defect-tolerance features \cite{Brandt2015,Zhang1998,Zakutayev2014}.

Despite these encouraging prospects, experimental PCEs of MChX solar cells remain below 10\%, far from their ideal Shockley-Queisser limit 
obtained from detailed balance (e.g., 30\% in BiSeI) \cite{Shockley1961,Miliaieva2025}. Understanding the origin of this performance decline is 
therefore essential for realizing their full technological potential and enabling large-scale deployment. While efficiency losses can arise from 
synthesis routes, morphology, or even device architecture, recent studies indicate that such extrinsic factors are not the primary bottlenecks in 
MChX absorbers \cite{Cano2025,Nielsen2025}. Instead, intrinsic defects are increasingly recognized as the decisive factor. Extended, high-dimensional 
defects such as grain boundaries, dislocations, and precipitates, although typically detrimental in many semiconductors, are found to be noncritical 
in MChX \cite{Cano2023} since their quasi-one-dimensional chain-like framework appears to hinder charge recombination 
\cite{Zhou2015,McKenna2021,Williams2020}. On the other hand, point defects, often overlooked in early assessments, may dominate nonradiative 
recombination and ultimately constrain device performance \cite{Savory2019,Lopez2025}.

Recent theoretical studies on BiSeI \cite{Lopez2025} and their binary counterparts Sb\textsubscript{2}S\textsubscript{3} \cite{Wang2025} and 
Sb\textsubscript{2}Se\textsubscript{3} \cite{Wang2024} have identified chalcogen vacancies as dominant recombination centers, drastically reducing 
their maximum PCE. Experimental works have also reported similar defect behaviour in related materials \cite{Hoye2022}. Consequently, sulfur and 
selenium vacancies (V\textsubscript{S} and V\textsubscript{Se}) might play a significant detrimental role in the broader MChX family. Yet, their 
relative impact on device performance may differ substantially depending on the interplay of the local chalcogen--halogen environment, which 
motivates a systematic, cross-material computational investigation of point defects on pnictogen chalcohalides.

Systematically estimating point-defect behaviour (e.g., formation energies and carrier capture coefficients) in MChX is highly nontrivial 
\cite{Alkauskas2014,Hoang2018}. First-principles calculations of defect formation energies and carrier capture processes \cite{Alkauskas2016} are 
computationally demanding and highly sensitive to methodological choices \cite{Mosquera-Lois2023}. Comparisons across different studies are often 
complicated \cite{Arrigoni2020,Lopez2023,Lopez2024-1} by variations in input parameters or approximations, often relying on error cancellation 
\cite{Lejaeghere2013}, which hinders finding any general trends. By contrast, a consistent, family-wide study performed with identical computational 
protocols offers a robust framework to disentangle chemical effects from methodological artifacts (Fig.~\ref{fig1}b) \cite{Lopez2025,Kavanagh2021}.

Here, we present a comprehensive investigation of the defect chemistry of four representative pnictogen chalcohalides (BiSI, BiSeI, BiSBr, and BiSeBr) 
using advanced first-principles density functional theory (DFT) calculations. A systematic comparison of all of them allows identifying selenium 
vacancies as the most detrimental defects within this family; in particular, they introduce deep electronic transition states near the Fermi level 
and enhance electron--phonon coupling, thus promoting nonradiative recombination in BiSeI and BiSeBr. In contrast, sulfur vacancies in BiSI and BiSBr 
are comparatively benign, owing to weaker electron-phonon coupling, despite presenting similar formation energies. This striking contrast originates 
from the larger ionic radius and higher polarizability of selenium atoms, which enhance lattice distortions around V\textsubscript{Se}, suggesting 
that the chemical environment (specifically, the interplay between chalcogen and halogen anions) dictates charge recombination activity in MChX. 
This work establishes design principles for defect engineering in pnictogen chalcohalides, while also offering a generalizable framework applicable 
to other emergent photovoltaic materials.

\section{Results}
\label{sec:results}
MChX crystallize into an orthorhombic phase (space group $Pnma$) characterized by one-dimensional columns held together by weak van der Waals forces 
(Fig.~\ref{fig1}a) \cite{Cano2025}, closely resembling the structure of binary pnictogen chalcogenides (Sb\textsubscript{2}S\textsubscript{3}, 
Sb\textsubscript{2}Se\textsubscript{3} and Bi\textsubscript{2}Se\textsubscript{3} \cite{Zhou2015}). Our DFT geometry optimizations for MChX yield 
lattice parameters that are in very good agreement with the available experimental data \cite{Lopez2024,Cano2025} (Supplementary Table~I). According 
to previous first-principles calculations and experimental observations \cite{Cano2023,Li2024}, MChX are thermodynamically stable against phase 
separation into secondary phases (Supplementary Fig.~1) and present indirect bandgaps in the range of 1 to 2~eV (Supplementary Fig.~2).

\subsection{Defect Formation Energies}
\label{subsec:defects}
Point defects play a pivotal role in the optoelectronic performance of semiconductors. They can compensate for doping-induced electrical imbalance, 
shift the Fermi energy level ($E_{\rm F}$), and act as deep nonradiative charge recombination centers that quench carrier lifetimes and limit 
photovoltaic performance. To assess their impact on MChX, we computed the formation energies ($E_{\rm f}$) of all relevant point defects as a 
function of the Fermi energy level (delimited by the valence band maximum and conduction band minimum, VBM~$\le E_F \le$~CBM) for BiSI, BiSeI, 
BiSBr and BiSeBr. In practice, low (high) defect formation energies imply high (low) equilibrium concentration of defects, as follows from the 
corresponding Boltzmann-like distribution.

\begin{figure*}[t]
    \centering
        \includegraphics[width=0.8\textwidth]{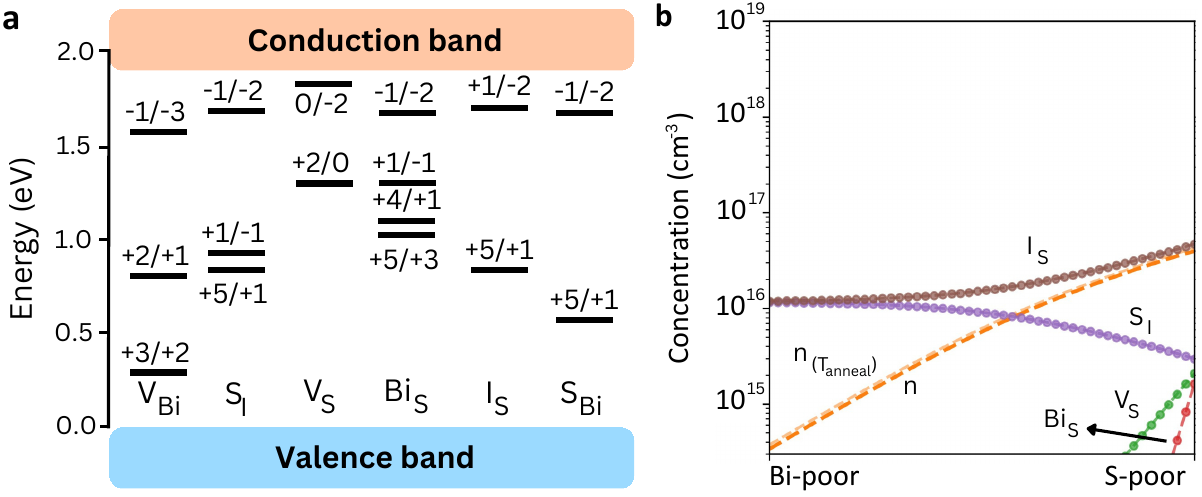}
        \caption{\textbf{Point-defect chemistry in BiSI.} \textbf{a.}~Charge-state transition levels of the most prominent (i.e., lowest energy)
        point defects determined for BiSI. \textbf{b.}~Defect concentrations of BiSI considering an annealing temperature of $550$~K. The chemical
        stability region is delimited by S-poor ($\mu$\textsubscript{Bi}, $\mu$\textsubscript{S}, $\mu$\textsubscript{I}) = (0, -0.58, -0.75)~eV
        and Bi-poor conditions ($\mu$\textsubscript{Bi}, $\mu$\textsubscript{S}, $\mu$\textsubscript{I}) = (-0.87, 0, -0.46)~eV.}
        \label{fig2}
\end{figure*}

Defect formation energy diagrams (Supplementary Figs.~3--11) provide a compact map of thermodynamic stability: the slope of each line reflects
the defect charge state, the intersections mark charge-state transition levels (acceptance or donation of electrons), and the relative position
of the curves identifies which defects are most favorable under given synthesis conditions. In terms of PV performance, the most detrimental
defects are those with low formation energies near the self-consistent Fermi level, $E^{\rm sc}_F$ \cite{Ganose2018,Freysoldt2014}. This
equilibrium Fermi level ensures charge neutrality across the defect and carrier populations in the system \cite{Squires2023}, and it can be
shifted with temperature and growth conditions.

The influence of chemical environment on $E_{\rm f}$ enters explicitly through the atomic chemical potentials (Methods), which bound the accessible 
growth window during synthesis. For MChX, the two pnictogen-poor and chalcogen-poor limits are particularly relevant (Table~I). These limits 
capture the experimentally relevant conditions where defect chemistry is expected to differ most strongly, while intermediate states interpolate 
between them. During experimental synthesis, chalcogen-poor conditions are frequently encountered owing to the high volatility of Se and S atoms 
at elevated temperatures \cite{Cano2023,Li2024}.

In the following sections, we focus on vacancies and antisites since these point defects consistently emerge as the lowest-energy across MChX.
In fact, interstitials exihibit significantly higher formation energies \cite{Lopez2025} (Supplementary Fig.~3), in agreement with experimental
observations in binary chalcogenides \cite{Lian2021,Mavlonov2020,Ma2019,Wen2018}, and are therefore excluded from our next detailed analysis.

\begin{table}[t]
    \centering
    \begin{tabular}{l|ccc|ccc}
        Material &  & Bi-poor &  &  & S/Se-poor &  \\ \hline
         & $\mu_{\rm Bi}$ & $\mu_{\rm S/Se}$ & $\mu_{\rm I/Br}$ & $\mu_{\rm Bi}$ & $\mu_{\rm S/Se}$ & $\mu_{\rm I/Br}$ \\
        BiSI   & -0.87 & 0 & -0.46 & 0 & -0.58 & -0.75 \\
        BiSeI  & -0.97 & 0 & -0.42 & 0 & -0.65 & -0.75 \\
        BiSBr  & -0.90 & 0 & -0.67 & 0 & -0.60 & -0.97 \\
        BiSeBr & -1.03 & 0 & -0.63 & 0 & -0.68 & -0.97
    \end{tabular}
    \caption{\textbf{Calculated chemical potential limits for the thermodynamic stability window of MChX compounds.} Bi-poor and S/Se-poor 
	conditions are the two limiting cases. Chemical potentials are expressed in units of eV.}
    \label{Table: synthesis conditions}
\end{table}

\subsubsection{BiSI}
\label{subsec:bisi}
At room temperature, $E^{\rm sc}_F$ lies $1.57$~eV above the VBM in Bi-poor environments and $1.70$~eV in S-poor environments, considering 
defect populations set at a realistic annealing temperature of $550$~K (Supplementary Fig.~4). The calculated $E^{\rm sc}_F$ exhibits a 
weak dependence on temperature (Supplementary Fig.~5), which denotes robust n-type character.

BiSI exhibits a high density of deep midgap charge-transition states that are likely to act as recombination centers (Fig.~\ref{fig2}a). 
These mostly include antisites, with Bi\textsubscript{S} showing three transitions (+5/+3, +3/+1, +1/-1) at $1.06$--$1.23$~eV from the VBM, 
complemented by Bi\textsubscript{I} (+5/0) and S\textsubscript{Bi} (+1/-1) at $1.14$ and $1.07$~eV, respectively. Sulfur vacancies with 
charge-transition state (+2/0) represent deep levels, positioned at $1.23$~eV from the VBM. 

Under S-poor conditions, V\textsubscript{S} exhibits a formation energy of $0.77$~eV at the self-consistent Fermi level, $E^{\rm sc}_F$ 
(Supplementary Fig.~5). Several antisite defects present also very low formation energies at $E^{\rm sc}_F$, the most critical cases being 
S\textsubscript{I} ($0.51$~eV) and Bi\textsubscript{S} ($0.54$~eV). Bi-poor synthesis conditions generally result in higher formation 
energies. For example, the formation energy of V\textsubscript{S} increases to $1.34$~eV. However, certain antisite defects still exhibit 
very low $E_{\rm f}$: $0.35$~eV for S\textsubscript{I} and $0.80$~eV for S\textsubscript{Bi}.

These results have a direct effect on the equilibrium defect populations, which are highest under S-poor conditions with, for example, 
$4.7 \times 10^{16}$~cm\textsuperscript{-3} for I\textsubscript{S}, $3.0 \times 10^{15}$~cm\textsuperscript{-3} for S\textsubscript{I}, 
$2.1 \times 10^{15}$~cm\textsuperscript{-3} for V\textsubscript{S} and $1.6 \times 10^{15}$~cm\textsuperscript{-3} for Bi\textsubscript{S} 
(Fig.~\ref{fig2}b). On the contrary, these populations are moderate under Bi-poor conditions with, for example, $1.2 \times 
10^{16}$~cm\textsuperscript{-3} for I\textsubscript{S} and S\textsubscript{I}.

\begin{figure*}[t]
    \centering
        \includegraphics[width=0.8\textwidth]{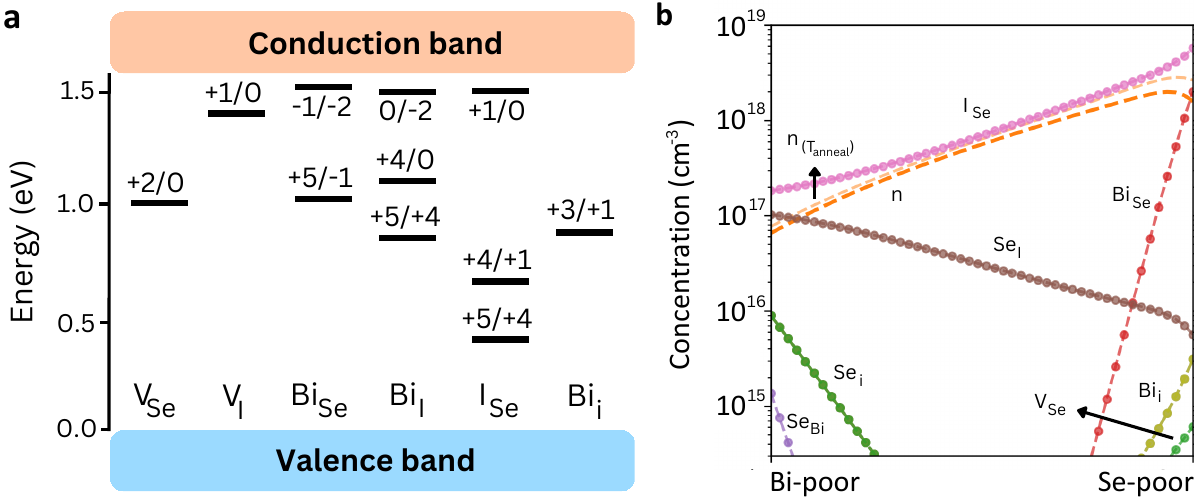}
        \caption{\textbf{Point-defect chemistry in BiSeI.} \textbf{a.}~Charge-state transition levels of the most prominent (i.e., 
lowest energy) point defects determined for BiSeI. \textbf{b.}~Defect concentrations of BiSeI considering an annealing temperature of 
$550$~K. The chemical stability region is delimited by Se-poor ($\mu$\textsubscript{Bi}, $\mu$\textsubscript{Se}, $\mu$\textsubscript{I}) 
= (0, -0.65, -0.75)~eV and Bi-poor conditions ($\mu$\textsubscript{Bi}, $\mu$\textsubscript{Se}, $\mu$\textsubscript{I}) = 
(-0.97, 0, -0.42)~eV.}
        \label{fig3}
\end{figure*}

\subsubsection{BiSeI}
\label{subsec:bisei}
In BiSeI, the defect landscape is similar to that of BiSI but with slightly different energetics (Supplementary Fig.~6). V\textsubscript{Se} 
forms at $0.82$~eV under Se-poor and $1.47$~eV under Bi-poor conditions. Bi\textsubscript{Se} is highly favorable, requiring only $0.34$~eV 
under Se-poor growth but increasing to $2.04$~eV in Bi-poor environments. The equilibrium Fermi level is again pinned high in the gap, 
$1.38$~eV above the VBM under Bi-poor and $1.46$~eV under Se-poor conditions, with little dependence on annealing temperature 
(Supplementary Fig.~7).

Similarly to BiSI, BiSeI exhibits deep levels such as Bi\textsubscript{Se} (+5/+1) positioned at $1.10$~eV, Bi\textsubscript{I} (+4/0) at 
$1.12$~eV from the VBM, Se\textsubscript{I} (+1/-1) at $0.87$~eV, Se\textsubscript{Bi} (+1/-1) at $1.06$~eV, and V\textsubscript{Se} (+2/0) 
at $1.08$~eV (Fig.~\ref{fig3}a). While in BiSeI there are fewer deep charge-transition states than in BiSI, the parallels between the two 
iodides indicate a shared tendency for potentially recombination-active centers.

The most critical antisites (Supplementary Fig.~6) are Bi\textsubscript{Se} ($0.34$~eV) and I\textsubscript{Se} ($0.43$~eV) under Se-poor 
conditions, while under Bi-poor conditions are I\textsubscript{Se} ($0.68$~eV) and Se\textsubscript{I} ($0.38$~eV). As in BiSI, the formation
 energy of vacancies are slightly higher than those of antisites. However, V\textsubscript{Se} emerges as a key recombination-active defect 
with $E_{\rm f} = 0.82$~eV, this formation energy being slightly lower than that of V\textsubscript{S} in BiSI.

Once again, we found that the equilibrium defect populations are highest under Se-poor conditions, with 
$5.8 \times 10^{18}$~cm\textsuperscript{-3} for I\textsubscript{Se}, $5.6 \times 10^{15}$~cm\textsuperscript{-3} for Se\textsubscript{I}, 
$6.1 \times 10^{14}$~cm\textsuperscript{-3} for V\textsubscript{Se} and $2.0 \times 10^{18}$~cm\textsuperscript{-3} for Bi\textsubscript{Se} 
(Fig.~\ref{fig3}b). Meanwhile, under Bi-poor conditions, we obtain $1.8 \times 10^{17}$~cm\textsuperscript{-3} for I\textsubscript{Se} and 
$1.0 \times 10^{17}$~cm\textsuperscript{-3} for Se\textsubscript{I}. These equilibrium defect concentrations in general are higher than 
found in BiSI. Consequently, it can be inferred that in BiSeI trap-mediated recombination would probably be more pronounced.

\begin{figure*}[t]
    \centering
        \includegraphics[width=0.8\textwidth]{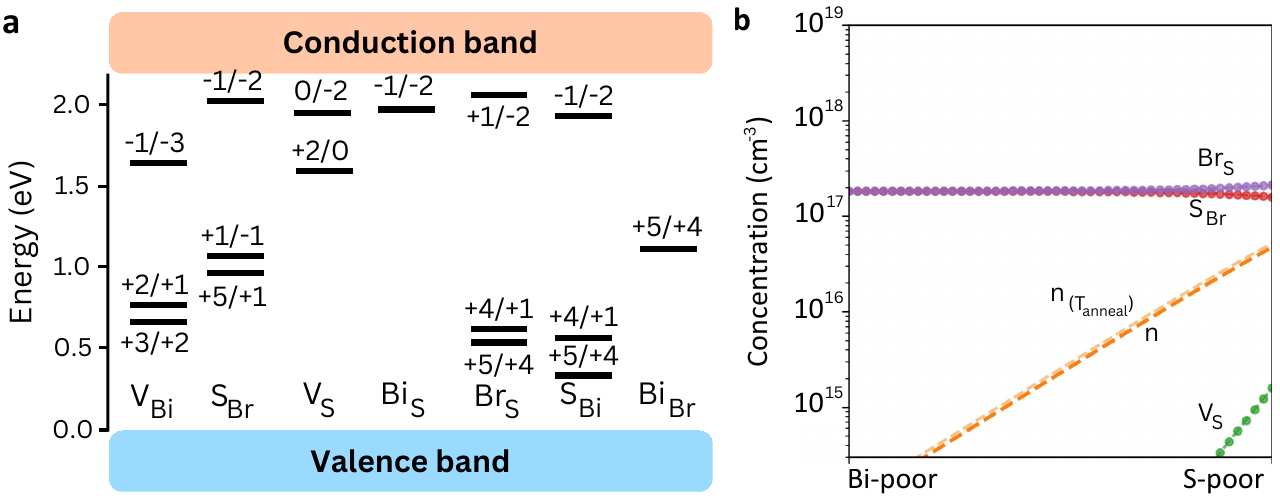}
        \caption{\textbf{Point-defect chemistry in BiSBr.} \textbf{a.}~Charge-state transition levels of the most prominent (i.e., lowest 
energy) point defects determined for BiSBr. \textbf{b.}~Defect concentrations of BiSBr considering an annealing temperature of $550$~K. 
The chemical stability region is delimited by S-poor ($\mu$\textsubscript{Bi}, $\mu$\textsubscript{S}, $\mu$\textsubscript{Br}) = 
(0, -0.60, -0.97)~eV and Bi-poor conditions ($\mu$\textsubscript{Bi}, $\mu$\textsubscript{S}, $\mu$\textsubscript{Br}) = 
(-0.90, 0, -0.67)~eV.}
        \label{fig4}
\end{figure*}

\subsubsection{BiSBr}
\label{subsec:bisbr}
BiSBr (Supplementary Fig.~8) shows the strongest n-type tendency since $E^{\rm sc}_F$ is pinned at $1.82$~eV above the VBM under Bi-poor 
and shifts further to $1.98$~eV in S-poor conditions (Supplementary Fig.~9). Deep levels arise from both cationic and anionic sites 
(Fig.~\ref{fig4}a): Bi\textsubscript{S} (+3/-1) at $1.34$~eV from the VBM, Bi\textsubscript{Br} (+4/+2, +2/0) at 1.25 and $1.47$~eV, 
respectively, S\textsubscript{Bi} (+1/-1) at $1.35$~eV, V\textsubscript{S} (+2/0) at $1.50$~eV, and V\textsubscript{Bi} (+1/-1) at 
$1.38$~eV. These states reduce the dominance of antisite deep transitions while introducing an additional recombination pathway through 
the Bi vacancy.

In BiSBr, the lowest-energy antisites are S\textsubscript{Br} and Br\textsubscript{S}, with $E_{\rm f}$ = 0.34~eV and $0.72$~eV under S-poor 
growth, respectively. Under Bi-poor conditions, S\textsubscript{Br} becomes particularly easy to form, with $E_{\rm f}$ dropping to $0.20$~eV,
 while S\textsubscript{Bi} also remains competitive ($0.73$~eV). Although vacancies formation is energetically less favorable, 
V\textsubscript{S} again emerges as the lowest-energy case, with $E_{\rm f}$ = 0.79~eV under S-poor, closely resembling previous cases. This 
hierarchy reflects the structural penalty for removing atoms from the quasi-1D columns, in contrast to antisites, which require minimal 
rearrangement.

The equilibrium defect populations were found to closely resemble those in BiSI, with $2.1 \times 10^{17}$~cm\textsuperscript{-3} for 
Br\textsubscript{S}, $1.6 \times 10^{17}$~cm\textsuperscript{-3} for S\textsubscript{Br}, and $1.6 \times 10^{15}$~cm\textsuperscript{-3} 
for V\textsubscript{S} in S-poor conditions, and $1.8 \times 10^{17}$~cm\textsuperscript{-3} for Br\textsubscript{S} and S\textsubscript{Br} 
in Bi-poor conditions (Fig.~\ref{fig4}b). As a result, it would be expected for BiSBr to have similar trap-mediated efficiency losses to 
BiSI.

\begin{figure*}[t]
    \centering
        \includegraphics[width=0.8\textwidth]{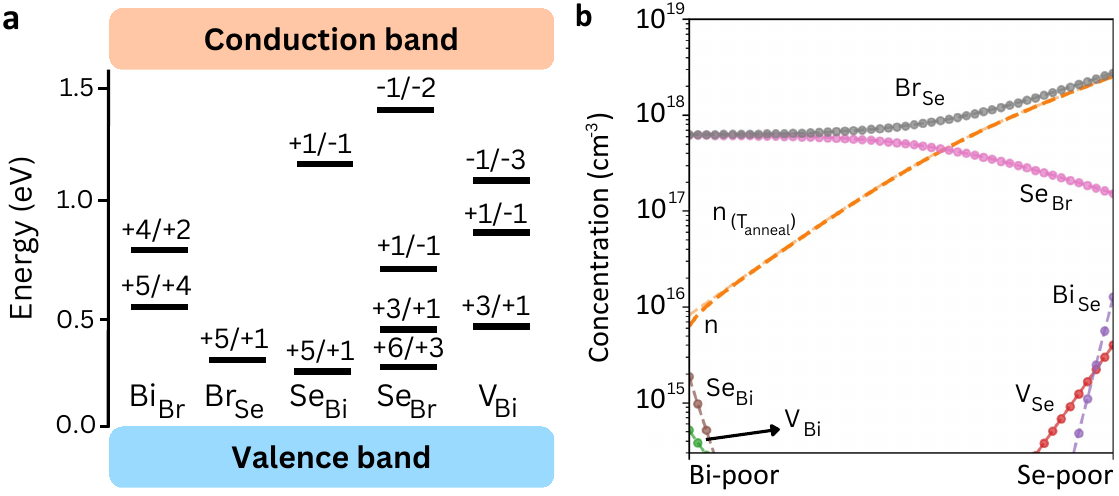}
        \caption{\textbf{Point-defect chemistry in BiSeBr.} \textbf{a.}~Charge-state transition levels of the most prominent (i.e., 
lowest energy) point defects determined for BiSeBr. \textbf{b.}~Defect concentrations of BiSeBr considering an annealing temperature of 
$550$~K. The chemical stability region is delimited by Se-poor ($\mu$\textsubscript{Bi}, $\mu$\textsubscript{Se}, $\mu$\textsubscript{Br}) 
= (0, -0.68, -0.97)~eV and Bi-poor conditions ($\mu$\textsubscript{Bi}, $\mu$\textsubscript{Se}, $\mu$\textsubscript{Br}) = (-1.03, 0, 
-0.63)~eV.}
        \label{fig5}
\end{figure*}

\subsubsection{BiSeBr}
\label{subsec:bisebr}
BiSeBr displays the richest set of midgap transitions (Fig.~\ref{fig5}a), including Bi\textsubscript{Se} (+5/-1) at $0.94$~eV from the VBM, 
Bi\textsubscript{Br} (+4/+2, +2/0) at $0.76$ and $1.14$~eV, respectively, Br\textsubscript{Bi} (0/-2) at $0.82$~eV, Se\textsubscript{Bi} 
(+1/-1) at $1.10$~eV, V\textsubscript{Se} (+2/0) at $1.08$~eV, and V\textsubscript{Bi} (+1/-1) at $0.99$~eV. This diversity exceeds that 
of BiSBr and the iodides, marking BiSeBr as the most defect-sensitive compound in the series.

In BiSeBr, Se-related defects dominate (Supplementary Fig.~10). V\textsubscript{Se} forms at $0.74$~eV under Se-poor conditions and 
$1.42$~eV under Bi-poor. Bi\textsubscript{Se} is also competitive under Se-poor conditions, with $E_{\rm f}$ = 0.56~eV, but not so 
under Bi-poor growth since its formation energy then increases to $2.43$~eV. The equilibrium Fermi level lies $1.33$~eV above the VBM 
under Bi-poor conditions and $1.49$~eV in Se-poor, indicating once again a robust n-type character (Supplementary Fig.~11).

The most critical antisites in Se-poor environments are Se\textsubscript{Br} ($0.44$~eV) and Br\textsubscript{Se} ($0.49$~eV) 
(Supplementary Fig.~10). Under Bi-poor conditions, Se\textsubscript{Br} ($0.26$~eV) and Se\textsubscript{Bi} ($0.59$~eV) remain highly 
favorable. Vacancies, though less favorable than antisites, still yield relatively low formation energies, with V\textsubscript{Se} 
($0.74$~eV) acting as a likely nonradiative recombination center, quite close in energy to V\textsubscript{S} in BiSI.

Again, the equilibrium defect populations are elevated under Se-poor conditions, with $2.8 \times 10^{18}$~cm\textsuperscript{-3} for 
Br\textsubscript{Se}, $1.5 \times 10^{17}$~cm\textsuperscript{-3} for Se\textsubscript{Br}, $1.3 \times 10^{16}$~cm\textsuperscript{-3} 
for Se\textsubscript{Br} and $4.0 \times 10^{15}$~cm\textsuperscript{-3} for V\textsubscript{Se} (Fig.~\ref{fig5}b). Under Bi-poor 
conditions, however, these equilibrium concentrations noticeably decrease: $6.2 \times 10^{17}$~cm\textsuperscript{-3} for 
Br\textsubscript{Se} and Se\textsubscript{Br}, $1.9 \times 10^{15}$~cm\textsuperscript{-3} for Se\textsubscript{Bi} and $5.2 \times 
10^{14}$~cm\textsuperscript{-3} for V\textsubscript{Bi}. These equilibrium defect concentrations are similar to those found in BiSeI, 
and slightly higher than for BiSBr, thus it can be expected that BiSeBr, along with BiSeI, present more trap-mediated recombination 
than BiSI and BiSBr.

\subsubsection{General trends}
\label{subsec:general}
Clear and systematic trends emerge from the defects formation results obtained for BiSI, BiSeI, BiSBr and BiSeBr 
(Figs.~\ref{fig2}--\ref{fig5}). Our calculations show that $E^{\rm sc}_F$ depends only weakly on annealing, indicating a robust n-type 
character across the MChX family. This behavior aligns with experimental observations that p-type doping is extremely challenging in 
these materials \cite{Cano2023}. Under chalcogen-poor conditions, Bi\textsubscript{Ch} antisites consistently rank among the lowest 
energy defects. This preference originates from the similar atomic environments found around Bi and chalcogen positions along the 
quasi-one-dimensional chains (Fig.~\ref{fig1}a). Likewise, chalcogen-on-halogen substitutions (Ch\textsubscript{X}) repeatedly show very 
low formation energies, as seen for both S- and Se-based compounds, which can be rationalized by the comparable ionic radii of these 
species (Fig.~\ref{fig1}b). While the formation of vacancies is in general energetically less favorable, V\textsubscript{Ch} consistently 
emerges as the dominant vacancy type. In Bi-poor environment, the formation of Ch\textsubscript{X} antisites remains the most favorable 
across all four materials, systematically followed by chalcogen-on-pnictogen (Ch\textsubscript{M}) and halogen-on-chalcogen 
(X\textsubscript{Ch}) substitutions. At the same time, V\textsubscript{M} becomes the vacancy species with the lowest formation energies.

These general defects formation energy trends can be intuitively understood in terms of the quasi-one dimensional crystal structure 
characteristic of MChX. Point defects are largely confined to the atomic columns in which they form, minimizing interactions with 
neighboring chains. Vacancies, which require substantial lattice relaxation within a column, exhibit moderate formation energies, whereas 
interstitials are even more energetically costly due to the extensive lattice disruption they introduce. By contrast, antisite defects, 
involving the replacement of one atom by another within the same column, require minimal structural rearrangement and are therefore easier 
to form. These patterns are consistent with prior observations in columnar pnictogen chalcogenides such as 
Sb\textsubscript{2}S\textsubscript{3} and Sb\textsubscript{2}Se\textsubscript{3} \cite{Savory2019,Wang2023}.

\subsection{Charge-Carrier Capture Coefficients}
\label{subsec:capture}
\begin{figure*}[t]
    \centering
        \includegraphics[width=\textwidth]{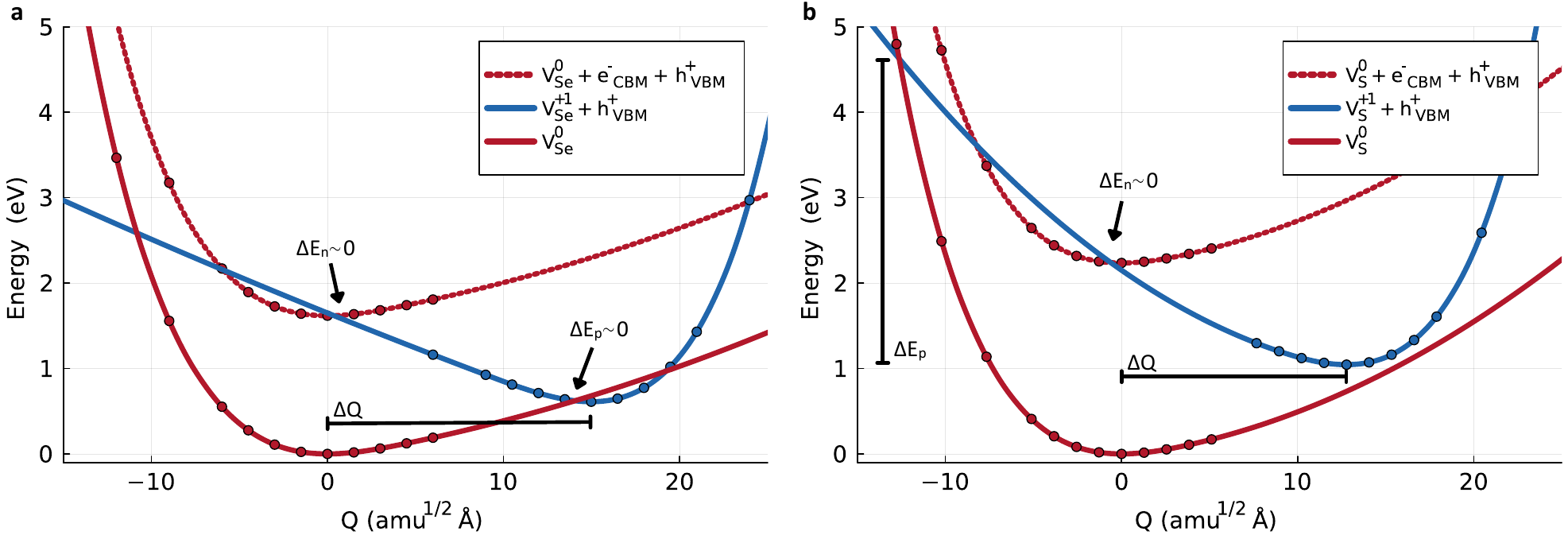}
        \caption{\textbf{Configuration coordinate diagrams for BiSeBr and BiSBr.} Configuration coordinate diagrams for \textbf{a.}~BiSeBr 
V\textsubscript{Se} (0/+1) and \textbf{b.}~BiSBr V\textsubscript{S} (0/+1). The dots represent potential energies computed from 
first-principles, and the solid lines are their corresponding quadratic spline interpolation and extrapolation. $\Delta Q$ represents the 
generalized distance between charge states and $\Delta E_x$ the electron ($n$)/hole ($p$) energy carrier capture barriers.}
        \label{fig6}
\end{figure*}

Carrier capture coefficients ($C_{n/p}$) provide a quantitative measure of the strength of electron–phonon coupling during defect-mediated 
charge exchange. While defect formation energies determine which defects are most abundant, carrier capture coefficients define which of 
those defects are electronically harmful by dictating how efficiently they can trap and recombine carriers \cite{Kim2020-2}. Defects with 
high $C_{n/p}$ act as efficient nonradiative recombination centers, drastically reducing carrier lifetimes and photovoltaic performance. 
Within multiphonon emission theory \cite{Alkauskas2014}, these coefficients are calculated from the overlap of vibrational wavefunctions 
corresponding to different charge states, whose equilibrium geometries are connected through a generalized configuration coordinate, 
$Q$ (Fig.~\ref{fig6}).

For BiSI (Supplementary Figs.~12,13), the most prominent recombination-active centers are associated with sulfur-related defects 
(Supplementary Table~II). We find particularly strong electron capture for the sulfur vacancy, with $C_{n}(\mathrm{V_{S}}, 0/+1) = 
2.94 \times 10^{-8}$~cm\textsuperscript{3}/s, accompanied by a notable contribution from the antisite transition $C_{n}(\mathrm{Bi_{S}},
 -1/0) = 2.40 \times 10^{-10}$~cm\textsuperscript{3}/s and $C_{n}(\mathrm{Bi_{S}}, +4/+5) = 1.06 \times 10^{-8}$~cm\textsuperscript{3}/s.

In BiSeI (Supplementary Figs.~14,15), the selenium vacancy emerges as the dominant nonradiative defect (Supplementary Table~III), 
exhibiting multiple strong capture channels: $C_{p}(\mathrm{V_{Se}}, 0/+1) = 5.89 \times 10^{-8}$~cm\textsuperscript{3}/s, 
$C_{n}(\mathrm{V_{Se}}, -1/0) = 6.71 \times 10^{-9}$~cm\textsuperscript{3}/s, and $C_{n}(\mathrm{V_{Se}}, +1/+2) = 1.24 \times 
10^{-9}$~cm\textsuperscript{3}/s. These values are approximately one order of magnitude higher than those observed in BiSI, indicating a 
more severe role of chalcogen vacancies in the selenides. Additional contributions arise from antisite defects such as Bi\textsubscript{Se}, 
with transitions $(-2/-1)$ and $(-1/0)$ yielding coefficients of $5.53 \times 10^{-10}$~cm\textsuperscript{3}/s and $5.88 \times 
10^{-10}$~cm\textsuperscript{3}/s, respectively. These antisites, while less active than $\mathrm{V_{Se}}$, still underscore the 
vulnerability of iodide compounds to midgap recombination centers.

In BiSBr (Supplementary Figs.~16,17), the sulfur vacancy again proves decisive (Supplementary Table~IV), with $C_{n}(\mathrm{V_{S}}, -2/-1) 
= 4.66 \times 10^{-9}$~cm\textsuperscript{3}/s. Interestingly, this material also displays significant antisite activity, with 
$C_{n}(\mathrm{Bi_{S}}, +1/+2) = 1.69 \times 10^{-9}$~cm\textsuperscript{3}/s and $C_{p}(\mathrm{Bi_{S}}, +3/+4) = 5.99 \times 
10^{-10}$~cm\textsuperscript{3}/s. Compared to the iodides, BiSBr distributes its recombination activity more evenly between vacancies and 
antisites, although presenting fewer recombination pathways.

BiSeBr (Supplementary Figs.~18,19) follows the same overall pattern as BiSeI, with selenium vacancies acting as the most harmful 
recombination centers (Supplementary Table~V). A very interesting result is the extraordinarily high hole capture rate $C_{p}(\mathrm{V_{Se}},
 0/+1) = 2.95 \times 10^{-5}$~cm\textsuperscript{3}/s, three orders of magnitude larger than in any other member of the series, firmly 
establishing $\mathrm{V_{Se}}$ as the dominant nonradiative defect in this compound. Additional strong recombination channels in BiSeBr 
include $C_{n}(\mathrm{V_{Se}}, -2/-1) = 1.31 \times 10^{-9}$~cm\textsuperscript{3}/s and the antisite transition $C_{p}(\mathrm{Bi_{Se}},
 -1/0) = 8.85 \times 10^{-9}$~cm\textsuperscript{3}/s.

The origin of the unusually large hole capture coefficient $C_p$ in BiSeBr is the almost zero hole energy capture barrier, $\Delta E_p$ 
(Fig.~\ref{fig6}a, Supplementary Table~V), which strongly promotes hole trapping. In fact, the relationship between the calculated potential 
energy surfaces and the resulting carrier-capture coefficients can be understood through the ``classical'' energy capture barriers 
\cite{Kavanagh2022,Alkauskas2014}. These barriers are defined as the energy difference between the equilibrium geometry of the initial 
charge state and the crossing point of the corresponding potential energy surfaces ($q \to q+1$ for $\Delta E_p$ and $q+1 \to q$ for $\Delta 
E_n$). In this framework, small barriers lead to fast carrier capture, whereas large barriers result in reduced capture coefficients. For 
example, BiSBr exhibits a much higher hole energy capture barrier than BiSeBr ($\Delta E_p = 3.58$~eV, Fig.~\ref{fig6}b and Supplementary 
Table~IV) resulting in a $C_p$ that is eight orders of magnitude smaller. Nevertheless, both compounds display comparable electron energy 
capture barriers, $\Delta E_n$, leading to similarly moderate electron capture coefficients.

These results reveal that chalcogen vacancies (particularly V\textsubscript{Se}) dominate the nonradiative recombination landscape in MChX, 
with capture coefficients orders of magnitude larger than those of antisites. These values are closely in line with results obtained for 
binary chalcogenides, which amount to $C_{n}(\mathrm{V_S}, +1/+2) = 7.64 \times 10^{-6}$~cm\textsuperscript{3}/s and $C_{p}(\mathrm{V_S}, 
+1/+2) = 2.25 \times 10^{-8}$~cm\textsuperscript{3}/s for Sb\textsubscript{2}S\textsubscript{3} \cite{Wang2025}, and $C_{n}(\mathrm{V_{Se}}, 
+1/+2) = 5.63 \times 10^{-6}$~cm\textsuperscript{3}/s and $C_{p}(\mathrm{V_{Se}}, +1/+2) = 1.22 \times 10^{-8}$~cm\textsuperscript{3}/s 
for Sb\textsubscript{2}Se\textsubscript{3} \cite{Wang2024}. For MChX, the sulfides exhibit somewhat weaker electron–phonon coupling than 
the selenides, explaining their lower recombination activity despite similar defect energetics. Antisites such as Bi\textsubscript{Ch}, 
although often abundant, generally display smaller capture coefficients and consequently can be regarded as comparatively benign.

\begin{table*}[t]
    \centering
    \begin{tabular}{l|c|ccc|ccc|ccc}
        \hline
        \multirow{3}{*}{Material} & \multirow{3}{*}{E\textsubscript{g} (eV)} 
        & \multicolumn{3}{c|}{Radiative limit} 
        & \multicolumn{6}{c}{Nonradiative limit} \\ \cline{6-11}
         &  & \multirow{2}{*}{$\eta$ (\%)} & \multirow{2}{*}{V\textsubscript{oc} (V)} & \multirow{2}{*}{FF (\%)} 
         & \multicolumn{3}{c|}{Bi-poor} & \multicolumn{3}{c}{Ch-poor} \\ %\cline{6-11}
         &  &  &  &  & $\eta$ (\%) & V\textsubscript{oc} (V) & FF (\%) & $\eta$ (\%) & V\textsubscript{oc} (V) & FF (\%) \\ \hline
        BiSI   & 1.96 & 23.63 & 1.66 & 92.08 & 23.63 & 1.66 & 92.08 & 23.20 & 1.66 & 90.66 \\
        BiSeI  & 1.60 & 30.38 & 1.33 & 90.54 & 30.38 & 1.33 & 90.54 & 24.23 & 1.08 & 88.81 \\
        BiSBr  & 2.24 & 17.65 & 1.92 & 92.94 & 17.65 & 1.92 & 92.94 & 17.26 & 1.91 & 91.25 \\
        BiSeBr & 1.62 & 27.96 & 1.35 & 90.66 & 27.96 & 1.35 & 90.66 & 18.21 & 0.91 & 87.23
    \end{tabular}
    \caption{\textbf{Estimated power conversion efficiencies and related properties considering different limits and growth conditions.} 
Band gap, power conversion efficiencies ($\eta$), open-circuit voltages (V\textsubscript{oc}) and fill factors (FF) for BiSI, BiSeI, 
BiSBr and BiSeBr estimated at the radiative and nonradiative limits.}
\end{table*}

\subsection{Power Conversion Efficiencies}
\label{subsec:pce}
\begin{figure}[t]
    \centering
        \includegraphics[width=0.5\textwidth]{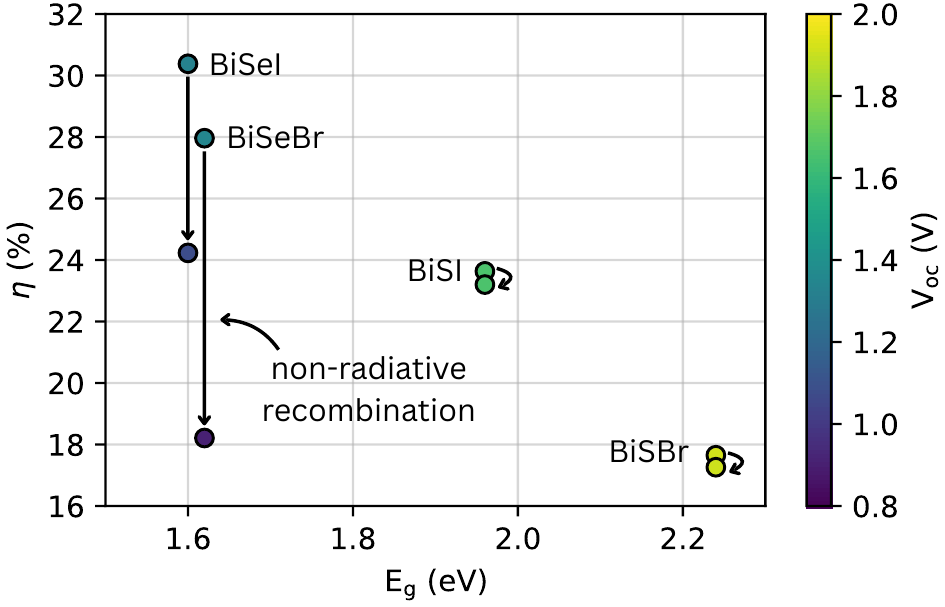}
        \caption{\textbf{Impact of non-radiative recombination in the estimated power conversion efficiencies and open-circuit voltage.} 
Power conversion efficiency and open circuit voltage of BiSI, BiSeI, BiSBr, and BiSeBr as a function of bandgap. Colored points indicate 
the open-circuit voltage for each material with a heatmap scale. Arrows indicate the drop in efficiency from the radiative to the 
nonradiative limit under chalcogen-poor conditions.}
	\label{fig7}
\end{figure}

The previous analysis of defect formation energies and charge-carrier capture coefficients highlights the potential impact of intrinsic 
defects on nonradiative recombination in MChX absorbers. To place these effects in context, we first consider the ideal case: the detailed 
balance model \cite{Shockley1961}, which neglects nonradiative carrier losses and establishes the theoretical radiative limits for the 
MChX family (Table~II and Fig.~\ref{fig7}).

At room temperature, and accounting for thickness-dependent absorptivity in a $700$~nm absorber layer, BiSI achieves a maximum power 
conversion efficiency ($\eta$) of 23.63\%, with an open-circuit voltage (V\textsubscript{oc}) of $1.66$~V and a fill factor (FF) of 
92.08\%. BiSeI, benefiting from its narrower band gap (Table~II), reaches an ideal $\eta$ of 30.38\% with V\textsubscript{oc} = 1.33~V 
and FF = 90.54\%, underscoring the intrinsic suitability of the iodides as outdoor photovoltaic absorbers. Among the bromides, BiSBr 
exhibits the lowest predicted efficiency, with a detailed balance $\eta$ of 17.65\%, despite a comparatively high V\textsubscript{oc} 
of $1.92$~V and FF of 92.94\%. Its poor efficiency arises directly from its wide band gap (Table~II), which severely limits photocurrent 
generation. BiSeBr, in contrast, performs on par with BiSeI, achieving a maximum $\eta$ of 27.96\% (V\textsubscript{oc} = 1.35~V, 
FF = 90.66\%). Taken together, these results demonstrate that S-based compounds inherently yield lower $\eta$ than Se-based compounds 
due to wider band gaps of the former.

Despite these favorable radiative limits, experimental $\eta$ values for MChX solar cells have not surpassed thus far 10\% 
\cite{Nie2021,Tiwari2019,Cano2023}. While extrinsic factors such as film morphology, interface quality, and device architecture undoubtedly 
contribute, our results indicate that defect-assisted nonradiative recombination is also a critical bottleneck. To quantify its impact, we 
evaluated the defect-limited maximum $\eta$ using the calculated carrier capture coefficients ($C_{n/p}$), defect concentrations, and 
related parameters (Methods) \cite{Dahan2013}.

Under Ch-poor synthesis conditions and an annealing temperature of 550~K, $\eta$ decreases only marginally for the sulfides, to 23.20\% 
for BiSI and 17.26\% for BiSBr, corresponding to a reduction of just 0.4\% compared to the radiative limit (Table~II). In contrast, for 
the selenides the effect is much stronger: $\eta$ drops to 24.23\% for BiSeI and 18.21\% for BiSeBr, representing losses of 6\% and 10\%, 
respectively (Fig.~\ref{fig7}). These reductions are accompanied by significant drops in V\textsubscript{oc}, falling to $1.08$~V for 
BiSeI and $0.91$~V for BiSeBr, while remaining nearly unchanged for BiSI and BiSBr. The fill factor is uniformly reduced by about 2\% 
across all Bi-based chalcohalides. By comparison, Bi-poor conditions have negligible influence, effectively minimizing the impact of 
trap-mediated recombination on photovoltaic performance (Table~II).

Separate $\eta$ calculations (Supplementary Table~VI), in which we isolate the recombination effects of chalcogen vacancies and 
Bi\textsubscript{Ch} antisites, reveal that the efficiency loss arises entirely from chalcogen vacancies, which act as strong 
nonradiative centers. Antisite defects, despite their low formation energies, exhibit extremely small carrier capture coefficients 
and can therefore be regarded as electronically benign. A similar picture holds across the broader family: in BiSI and BiSBr, 
sulfur-related defects (V\textsubscript{S}, Bi\textsubscript{S}) produce negligible recombination losses, keeping their efficiencies 
effectively identical to the radiative limit. Thus, the dominant factor limiting nonradiative performance in MChX is the presence of 
chalcogen-site vacancies, whereas antisite defects appear comparatively harmless for device operation.

\section{Discussion}
The nonradiative efficiency losses in MChX (Table~II and Fig.~\ref{fig7}) are comparable to those observed in other light absorber 
materials, such as Cu\textsubscript{2}ZnSnS\textsubscript{4} \cite{Kim2021}, Cu\textsubscript{2}ZnSnSe\textsubscript{4} \cite{Kim2020-2}, 
and CdTe \cite{Kavanagh2021}. Both BiSeI and BiSeBr exhibit pronounced reductions in PCE relative to their radiative limits of approximately 
6\% and 10\%, respectively. These losses are accompanied by marked decreases in open-circuit voltage (V\textsubscript{oc} = 1.08~V for 
BiSeI and 0.91~V for BiSeBr) and fill factor (88.81\% and 87.23\%, respectively), reflecting a deterioration in electronic quality under 
illumination and a reduced charge-transport efficiency. Interestingly, this loss mechanism arises not from exceptionally strong recombination
 activity but from the high equilibrium concentration of selenium vacancies (V\textsubscript{Se}), whose relatively low formation energies 
($\approx 0.8$~eV) lead to abundant deep-level traps. Hence, performance limitations in MChX absorbers are mainly driven by the prevalence of 
intrinsic point defects rather than their individual recombination strengths. Effective passivation of these defects is therefore a promising
 avenue for improving device performance. Our first-principles results demonstrate that adopting Bi-poor synthesis conditions can 
significantly raise defect formation energies, thereby reducing their population. Under such conditions, the calculated efficiency for 
BiSeI and BiSeBr are 30.38\% and 27.96\%, respectively, closely approaching their ideal detailed-balance limits.

A complementary route to mitigate these defect-induced PCE losses may involve targeted ion substitution, which provides a chemically 
guided approach to engineering defect tolerance. The similar van der Waals radii of Bi (r\textsubscript{Bi} = 207 pm), Se 
(r\textsubscript{Se} = 190 pm), and halogens such as Br (r\textsubscript{Br} = 185 pm) or I (r\textsubscript{I} = 198 pm) enable facile 
atomic interchange during synthesis, promoting antisite and vacancy formation. To counteract this atomic interchangeability, we propose 
controlled substitutions involving ions with larger size mismatches to hinder defect diffusion and reduce antisite formation, while 
maintaining desirable optoelectronic characteristics. This rationale explains the enhanced defect tolerance observed in the sulfides 
BiSI and BiSBr, where the substantial ionic radius contrast among Bi, S (r\textsubscript{S} = 180 pm), and halogens suppresses in part 
defect formation. These insights suggest that careful compositional tuning, particularly through I-on-Br substitution, offers a promising 
strategy to reduce formation of defect in MChX semiconductors and boost their efficiency.

The findings presented in this study are significant for the field of photovoltaics, as they provide a pathway to enhance the efficiency 
of MChX light absorbers, a promising class of nontoxic and thermodynamically stable materials. By addressing defect chemistry and proposing 
effective passivation strategies, this study not only bridges the gap between theoretical predictions and experimental performance but also 
establishes a framework for designing more efficient solar absorbers with improved defect tolerance. These insights could pave the way for 
next-generation photovoltaic technologies with higher efficiencies and broader applicability.

Based on our theoretical findings, we suggest two promising experimental MChX research directions: (i)~exploring ion substitution strategies 
(either full replacement or partial extrinsic doping) to suppress defect formation and weaken electron–phonon interactions that drive 
nonradiative recombination, and (ii)~pursuing synthesis under chalcogen-rich conditions, which would intrinsically minimize the 
concentration of detrimental vacancies and thereby reduce defect-mediated efficiency losses.

\section{Conclusions}
\label{sec:conclusions}
This work identifies intrinsic point defects as a key bottleneck limiting the photovoltaic performance of pnictogen chalcohalides. 
Through a systematic first-principles computational study involving BiSI, BiSeI, BiSBr, and BiSeBr, we show that chalcogen vacancies 
dominate nonradiative recombination, with selenium vacancies being particularly harmful due to their deep electronic transition states 
and strong electron–phonon coupling. In contrast, sulfur vacancies couple more weakly and are comparatively benign, highlighting the 
decisive influence of local chemical environment on defect tolerance.

Our defect-limited efficiency analysis shows that while sulfides retain efficiencies close to their radiative limits, the selenides 
suffer losses of up to 10\% on efficiency, accompanied by severe reductions in open-circuit voltage. Crucially, we also demonstrate 
that synthesis conditions and targeted anion substitutions can mitigate the impact of harmful vacancies: Bi-poor growth suppresses 
V\textsubscript{Se} formation, while S-for-Se or I-for-Br substitutions improve defect tolerance without sacrificing favorable 
optoelectronic properties. These findings establish a general design principles for defect passivation in emergent optoelectronic
materials, paving the way toward more efficient and defect-tolerant photovoltaic technologies.

\section*{Methods}
\subsection{First-Principles Calculations}
\textit{Ab initio} calculations based on density functional theory (DFT) were  performed to analyse point defects in MChX. These calculations were conducted with the \verb!VASP! software package \cite{Kresse1996} using the generalized gradient approximation to the exchange-correlation energy due to Perdew \textit{et al.} (PBE) \cite{Perdew2008}. Since MChX are van der Waals materials, long-range dispersion interactions were taken into account through the D3 scheme \cite{Grimme2010}. Spin–orbit coupling effects, which are particularly relevant for Bi-based MChX \cite{Ganose2018,Ganose2016,Lopez2025} (see details in Supplementary Discussion), were taken into account along with range-separated hybrid functionals containing an exact Hartree-Fock exchange fraction of $25$\% (i.e., HSE+SOC \cite{Schimka2011,Heyd2003,Krukau2006}). All defective atomic structures were fully optimized at the HSE+D3+SOC level, a methodology shown to accurately reproduce experimental results for MChX and other similar materials \cite{Lopez2024,Pan2018,Lopez2025}. The projector augmented-wave method was used to represent the ionic cores \cite{Blochl1994} and for each element the maximum possible number of valence electronic states was considered. Wave functions were represented in a plane-wave basis typically truncated at $300$~eV. By using these parameters and a dense $\Gamma$-centered \textbf{k}-point grid for reciprocal space Brillouin zone integration of $2 \times 1 \times 2$, the resulting energies were converged to within $1$~meV per formula unit. In the geometry relaxations, a tolerance of $0.5$~meV$\cdot$\AA$^{-1}$ was imposed in the atomic forces. Defect calculations were performed in $3 \times 2 \times 1$ ($12.6 \times 17.2 \times 10.3$~\AA~for BiSI, $12.8 \times 18.1 \times 11.3$~\AA~for BiSeI, $12.1 \times 16.1 \times 9.5$~\AA~for BiSBr and $12.2 \times 16.3 \times 10.1$~\AA~for BiSeBr) supercells.

\subsection{Exploration of the Potential Energy Surface}
Conventional approaches to generating defect configurations \cite{Freysoldt2014} from pristine cells fail to find many lowering-energy conformations, which might have a crucial effect in the conclusions. Therefore, once a defect is generated (e.g., extracting a bismuth atom), we look for distortions of the initial lattice configuration to locally explore the potential energy surface. These distortions were generated with the \verb!ShakeNBreak! software package \cite{Mosquera-Lois2022,Mosquera-Lois2023}.

Initially, all the trial defect configurations were relaxed using $\Gamma$-point reciprocal space sampling and the HSE+D3 functional. Only the minimum-energy configurations were kept. Next, the ionic relaxations were repeated considering a larger \textbf{k}-point grid of $2 \times 1 \times 2$ ($\Gamma$-centered). After that, the relaxations were repeated considering SOC corrections. Finally, single-point energy calculations were performed using the previously-converged electronic wavefunctions and equilibrium structures. Non-spherical contributions to the gradient of the density within the PAW spheres were taken into account to improve numerical accuracy.

\subsection{Point-Defect Formation Energies}
Point defects include vacancies, where an atom is removed from the lattice (e.g., V\textsubscript{Bi}), antisites, where an atom is replaced by another of a different species (e.g., I\textsubscript{S}), and interstitials, where an atom occupies a nonequilibrium lattice site (e.g., Br\textsubscript{i}). Computational approaches for studying these are well-established, relying on accurate first-principles energy calculations combined with exhaustive exploration of the defect local environment \cite{Mosquera-Lois2022,Freysoldt2009,Kumagai2014}. For this work, we employed the supercell approach, which involves modeling point defects within sufficiently large supercells to minimize spurious interactions. We systematically analyzed all possible vacancy and antisite defects, considering both neutral and charged states. The doped simulation package \cite{Kavanagh2024} was used to generate defect structures and calculation inputs, determine chemical potential limits, and analyze the defects simulation results. The \verb!ShakeNBreak! \cite{Mosquera-Lois2022,Mosquera-Lois2021} defect structure-search approach was employed, revealing numerous significant energy-lowering reconstructions, consistent with observations in similar low-dimensional chalcogenide systems \cite{Mosquera-Lois2024,Wang2025,Wang2023}.

The formation energy of a point defect with charge $q$, $D^{q}$, can be expressed as \cite{Freysoldt2009}:
\begin{equation}
    \begin{gathered}
        E_{\rm f}(D^q) = E_T(D^q) - E_T({\rm pristine}) + \\
         + q E_F + \sum_i n_i \mu_i + E_{corr} (D^q)~,
    \end{gathered}
    \label{Eq. defect formation energy}
\end{equation}
where $E_T(D^q)$ and $E_T({\rm pristine)}$ are the static energies of defected and pristine supercells (energy per formula unit), respectively, $\mu_i$ corresponds to chemical potential of species $i$ (this is, the energy required to extract one single atom), $n_i$ the number of  extracted atoms (positive or negative if extracted or added to the pristine cell, respectively), $E_F$ the Fermi energy (energy needed to extract an electron), and $E_{corr}$ the finite-size corrections based on spurious interactions between charged defects due to the periodic boundary conditions.

Here we considered two different contributions to the finite-size energy correction: point-charge (due to the spurious electrostatic interactions of a defect with its images) and band-alignment corrections (charged defects spuriously change the electrostatic potential of the system). Both corrections are computed together from an extension of the Freysoldt-Neugebauer-Van de Walle \cite{Freysoldt2009} correction scheme to anisotropic materials \cite{Kumagai2014}, as implemented in the \verb!doped! defect simulation package \cite{Kavanagh2024}.

\subsection{Defect-Limited Efficiency}
The strength of nonradiative recombination was quantified through the electron and hole capture coefficients corresponding to each charge state of the defect, as obtained with the \verb!CarrierCapture.jl! package \cite{Alkauskas2014,Kim2020}.

In this theoretical framework, the maximum photovoltaic efficiency limited by defects \cite{Dahan2013} for an absorber of thickness $W$, under an incident photon flux $\Phi$ at a photon energy $E$ and bias voltage $V$, is expressed as:
\begin{equation}
    \eta = \max_V \left( \frac{J V}{q \int_0^{\infty} E \Phi(E) dE} \right)~,
\end{equation}
where $q$ denotes the elementary charge and $J$ the maximum defect-limited current density, defined by:
\begin{equation}
    J (W, V) = J_{SC} (W) + J_0^{rad} (W, V) + J_0^{nonrad} (W, V)~,
\end{equation}
Here, $J_{SC}$ represents the short-circuit current, while $J_0^{rad}$ corresponds to the radiative saturation current. These two contributions describe recombination through photon emission, whereas $J_0^{nonrad}$ accounts for nonradiative processes:
\begin{equation}
    J_{SC} (W) = q \int_0^{\infty} a (E, W) \Phi (E) dE
\end{equation}
\begin{equation}
    \begin{gathered}
        J_0^{rad} (W, V) = q \frac{2 \pi}{c^2 h^3} \left( 1 - e^{\frac{q V}{k_B T}} \right) \times \\
        \times \int_0^{\infty} a (E, W) \left( e^{\frac{E}{k_B T}} - 1 \right)^{-1} E^2 dE
    \end{gathered}
\end{equation}

\begin{equation}
    J_0^{nonrad} (W, V) = - q W R^{SRH} (V)~,
\end{equation}
with $a$ the absorptivity (assuming unit quantum efficiency such that each absorbed photon produces an electron–hole pair). The Shockley–Read–Hall recombination rate is approximated as:
\begin{equation}
    R_{SRH} \approx \sum \Delta n/p ~N_T C_{n/p}~,
    \label{Eq. recombination rate}
\end{equation}
where $\Delta n$ and $\Delta p$ are the excess carrier concentrations, $N_T$ the defect concentration, and the summation runs over all recombination-active centers. The capture coefficient \cite{Kim2020-2,Alkauskas2014}, $C_{n/p}$, depends on the electron–phonon interaction strength ($W_{ct}$) and the overlap of vibrational wave functions ($\langle \zeta_{cm} | \Delta Q | \zeta_{tn} \rangle$):
\begin{equation}
    \begin{gathered}
        C_{n/p} = \Omega \frac{2 \pi}{\hbar} |W_{ct}|^2 \sum_{m,n} w_m \langle \zeta_{cm} | \Delta Q | \zeta_{tn} \rangle^2 \times \\
        \times \delta(\Delta E_{n/p} + \epsilon_{cm} - \epsilon_{tn})~,
    \end{gathered}
    \label{Eq. carrier capture}
\end{equation}
where $\Omega$ is the supercell volume, $g$ the defect degeneracy, $\zeta$ the phonon wave function, and $\Delta Q$ the effective configuration coordinate, with $c$ and $t$ labeling free-carrier and trap states, respectively. The temperature dependence arises from the thermal occupation probability $w_m$ of the initial vibrational mode. To incorporate the role of capture dynamics on photovoltaic efficiency, calculations were carried out with the \verb!TLC! code \cite{Kim2020-2}.
\\

\section*{Data availability}
The data that support the findings of this study have been made publicly available \cite{database}, comprising the single-point energy and local potential calculations for all relaxed defects, along with all input files needed for reproducibility using the \texttt{doped} Python package \cite{Kavanagh2024}.
\\

\section*{Acknowledgments}
C.L. acknowledges support from the Spanish Ministry of Science, Innovation and Universities under an FPU grant.
C.C. acknowledges support by MICIN/AEI/10.13039/501100011033 and ERDF/EU under the grants TED2021-130265B-C22,
TED2021-130265B-C21, PID2023-146623NB-I00 and PID2023-147469NB-C21 and by the Generalitat de Catalunya under the 
grants 2021SGR-00343, 2021SGR-01519 and 2021SGR-01411.
P.B. acknowledges support from the predoctoral program AGAUR-FI ajuts (2024 FI-100070) Joan Oró. The authors also 
thankfully acknowledge technical support for the computational resources at MareNostrum4 provided by the Barcelona 
Supercomputing Center (FI-2024-1-0025, FI-2024-1-0005, FI-2024-2-0006 and FI-2024-2-0003). 
E.S. acknowledges the European Union H2020 Framework Program SENSATE project: Low dimensional semiconductors for optically 
tuneable solar harvesters (grant Number 866018), the Spanish Ministry of Science and Innovation ACT-FAST (PCI2023-145971-2),
and the ICREA Academia program.
All authors gratefully acknowledge the Renew-PV European COST action (CA21148) for funding our collaboration.
This work is part of the Maria de Maeztu Units of Excellence Programme CEX2023-001300-M funded by MCIN/AEI (10.13039/501100011033). 
\\

\section*{Author contributions}
C.L., E.S. and C.C. conceived the study and planned the research, which was discussed in-depth with the rest of the co-authors. C.L. and S.K. performed the first-principles calculations and analyzed the results. The manuscript was written by C.L. and C.C., with substantial input from the rest of the co-authors.
\\

\section*{Additional information}
Supplementary information is available in the online version of the paper.
\\

\section*{Competing financial interests}
The authors declare no competing financial interests.
\\

\bibliographystyle{unsrt}

\begin{thebibliography}{10}

\bibitem{Hermann2006}
W.~Hermann.
\newblock Quantifying global exergy resources.
\newblock {\em Energy}, 31:1685, 2006.

\bibitem{Peter2011}
L.~M. Peter.
\newblock Towards sustainable photovoltaics: the search for new materials.
\newblock {\em Philos. Trans. R. Soc. A}, 369:1840--1856, 2011.

\bibitem{Lopez2024}
C.~López, I.~Caño, D.~Rovira, P.~Benítez, J.~M. Asensi, Z.~Jehl, J.-L.
  Tamarit, E.~Saucedo, and C.~Cazorla.
\newblock Machine-learning aided first-principles prediction of earth-abundant
  pnictogen chalcohalide solid solutions for solar-cell devices.
\newblock {\em Adv. Funct. Mater.}, 34:2406678, 2024.

\bibitem{Ganose2016}
A.~M. Ganose, K.~T. Butler, A.~Walsh, and D.~O. Scanlon.
\newblock Relativistic electronic structure and band alignment of bisi and
  bisei: candidate photovoltaic materials.
\newblock {\em J. Mater. Chem. A}, 4:2060--2068, 2016.

\bibitem{Nielsen2025}
R.~S. Nielsen, Á.~L. Álvarez, A.~G. Medaille, I.~Caño, A.~N. Güell, C.~L.
  Álvarez, C.~Cazorla, D.~R. Ferrer, Z.~J. Li-Kao, E.~Saucedo, and
  M.~Dimitrievska.
\newblock Parallel exploration of the optoelectronic properties of
  (sb,bi)(s,se)(br,i) chalcohalides.
\newblock {\em J. Mater. Chem. A}, 13:31727--31739, 2025.

\bibitem{Nie2019}
R.~Nie, J.~Im, and S.~I. Seok.
\newblock Efficient solar cells em- ploying light-harvesting sb0.67 bi0.33 si.
\newblock {\em Adv. Mater.}, 31:1808344, 2019.

\bibitem{Li2024}
S.~Li, Z.~Huang, Y.~Ding, C.~Zhang, J.~Yu, Q.~Feng, and J.~Feng.
\newblock Growth of bisbr microsheet arrays for enhanced photovoltaics
  performance.
\newblock {\em Small}, 20:2306964, 2024.

\bibitem{Cano2023}
I.~Caño, A.~Navarro, E.~Maggi, M.~Barrio, J.~L. Tamarit, S.~Svatek,
  E.~Antolín, S.~Yan, E.~Barrena, B.~Galiana, M.~Placidi, J.~Puigdollers, and
  Saucedo E.
\newblock Sbsei and sbsebr micro-columnar solar cells by a novel high
  pressure-based synthesis process.
\newblock {\em J. Mat. Chem. A}, 11:17616, 2023.

\bibitem{Cano2025}
I.~Caño, A.~Navarro-Güell, E.~Maggi, A.~Gon~Medaille, D.~Rovira,
  A.~Jimenez-Arguijo, O.~Segura, A.~Torrens, M.~Jimenez, C.~López,
  P.~Benítez, C.~Cazorla, Z.~Jehl, Y.~Gong, J.-M. Asensi, L.~Calvo-Barrio,
  L.~Soler, J.~Llorca, J.-L. Tamarit, B.~Galiana, M.~Dimitrievska,
  N.~Ruiz-Marín, H.~Z. Chun, L.~Wong, J.~Puigdollers, M.~Placidi, and
  E.~Saucedo.
\newblock Ribbons of light: Emerging (sb,bi)(s,se)(br,i) van der waals
  chalcohalides for next-generation energy applications.
\newblock {\em Small}, 21:e05430, 2025.

\bibitem{Brandt2015}
R.~E. Brandt, V.~Stevanovic, D.~S. Ginley, and T.~Buonassisi.
\newblock Identifying defect-tolerant semiconductors with high minority-carrier
  lifetimes: beyond hybrid lead halide perovskites.
\newblock {\em MRS Communications}, 5:265--275, 2015.

\bibitem{Zhang1998}
S.~B. Zhang, S.-H. Wei, A.~Zunger, and H.~Katayama-Yoshida.
\newblock Defect physics of the ${\mathrm{cuinse}}_{2}$ chalcopyrite
  semiconductor.
\newblock {\em Phys. Rev. B}, 57:9642--9656, 1998.

\bibitem{Zakutayev2014}
A.~Zakutayev, C.~M. Caskey, A.~N. Fioretti, D.~S. Ginley, J.~Vidal,
  V.~Stevanovic, E.~Tea, and S.~Lany.
\newblock Defect tolerant semiconductors for solar energy conversion.
\newblock {\em J. Phys. Chem. Lett.}, 5:1117--1125, 2014.

\bibitem{Shockley1961}
W.~Shockley and H.~J. Queisser.
\newblock Detailed balance limit of efficiency of p‐n junction solar cells.
\newblock {\em Journal of Applied Physics}, 32:510--519, 1961.

\bibitem{Miliaieva2025}
D.~Miliaieva, V.~Nadazdy, M.~Koltsov, C.~López, H.~Saeeyekta, J.~Kuliček,
  C.~Cazorla, E.~Saucedo, Grzibovskis R., and A.~Vembris.
\newblock Electronic structure and defect states in bismuth and antimony
  sulphides identified by energy-resolved electrochemical impedance
  spectroscopy.
\newblock {\em J. Phys. Energy}, 7:035012, 2025.

\bibitem{Zhou2015}
Y.~Zhou, L.~Wang, S.~Chen, and et~al.
\newblock Thin-film sb2se3 photovoltaics with oriented one-dimensional ribbons
  and benign grain boundaries.
\newblock {\em Nat. Photonics}, 9:409--415, 2015.

\bibitem{McKenna2021}
K.~P. McKenna.
\newblock Self-healing of broken bonds and deep gap states in sb2se3 and sb2s3.
\newblock {\em Adv. Electron. Mater.}, 7:2000908, 2021.

\bibitem{Williams2020}
R.~E. Williams, Q.~M. Ramasse, K.~P. McKenna, L.~J. Phillips, P.~J. Yates,
  O.~S. Hutter, K.~Durose, J.~D. Major, and B.~G. Mendis.
\newblock Evidence for self-healing benign grain boundaries and a highly
  defective sb2se3-cds interfacial layer in sb2se3 thin-film photovoltaics.
\newblock {\em ACS Appl. Mater. Interfaces}, 12:21730–21738, 2020.

\bibitem{Savory2019}
C.~N. Savory and D.~O. Scanlon.
\newblock The complex defect chemistry of antimony selenide.
\newblock {\em J. Mater. Chem. A}, 7:10739--10744, 2019.

\bibitem{Lopez2025}
C.~López, S.~R. Kavanagh, P.~Benítez, E.~Saucedo, A.~Walsh, David~O. Scanlon,
  and C.~Cazorla.
\newblock Chalcogen vacancies rule charge recombination in pnictogen
  chalcohalide solar-cell absorbers.
\newblock {\em ACS Energy Lett.}, 10:3562--3569, 2025.

\bibitem{Wang2025}
X.~Wang, S.~R. Kavanagh, and A.~Walsh.
\newblock Sulfur vacancies limit the open-circuit voltage of sb2s3 solar cells.
\newblock {\em ACS Energy Lett.}, 10:161–167, 2025.

\bibitem{Wang2024}
X.~Wang, S.~R. Kavanagh, D.~O. Scanlon, and A.~Walsh.
\newblock Upper efficiency limit of sb2se3 solar cells.
\newblock {\em Joule}, 8:2105--2122, 2024.

\bibitem{Hoye2022}
R.~L.~Z. Hoye, J.~Hidalgo, R.~A. Jagt, J.-P. Correa-Baena, T.~Fix, and J.~L.
  MacManus-Driscoll.
\newblock The role of dimensionality on the optoelectronic properties of oxide
  and halide perovskites, and their halide derivatives.
\newblock {\em Adv. Energy Mater.}, 12:2100499, 2022.

\bibitem{Alkauskas2014}
A.~Alkauskas, Q.~Yan, and C.~G. Van~de Walle.
\newblock First-principles theory of nonradiative carrier capture via
  multiphonon emission.
\newblock {\em Phys. Rev. B}, 90:075202, 2014.

\bibitem{Hoang2018}
K.~Hoang and M.~D. Johannes.
\newblock Defect physics in complex energy materials.
\newblock {\em J. Phys.: Condens. Matter}, 30:293001, 2018.

\bibitem{Alkauskas2016}
A.~Alkauskas, M.~D. McCluskey, and C.~G. Van~de Walle.
\newblock Tutorial: Defects in semiconductors—combining experiment and
  theory.
\newblock {\em Journal of Applied Physics}, 119:181101, 2016.

\bibitem{Mosquera-Lois2023}
I.~Mosquera-Lois, S.~R. Kavanagh, A.~Walsh, and D.~O. Scanlon.
\newblock Identifying the ground state structures of point defects in solids.
\newblock {\em npj Comput. Mater.}, 9:25, 2023.

\bibitem{Arrigoni2020}
M.~Arrigoni and G.~K.~H. Madsen.
\newblock A comparative first-principles investigation on the defect chemistry
  of tio2 anatase.
\newblock {\em The Journal of Chemical Physics}, 152:044110, 2020.

\bibitem{Lopez2023}
C.~López, A.~Emperador, E.~Saucedo, R.~Rurali, and C.~Cazorla.
\newblock Universal ion-transport descriptors and classes of inorganic
  solid-state electrolytes.
\newblock {\em Mater. Horiz.}, 10:1757, 2023.

\bibitem{Lopez2024-1}
C.~López, R.~Rurali, and C.~Cazorla.
\newblock How concerted are ionic hops in inorganic solid-state electrolytes?
\newblock {\em J. Am. Chem. Soc.}, 146:8269--8279, 2024.

\bibitem{Lejaeghere2013}
K.~Lejaeghere, V.~Van~Speybroeck, G.~Van~Oost, and S.~Cottenier.
\newblock Error estimates for solid-state density-functional theory
  predictions: An overview by means of the ground-state elemental crystals.
\newblock {\em Critical Reviews in Solid State and Materials Sciences},
  39:1–24, 2013.

\bibitem{Kavanagh2021}
S.~R. Kavanagh, A.~Walsh, and D.~O. Scanlon.
\newblock Rapid recombination by cadmium vacancies in cdte.
\newblock {\em ACS Energy Lett.}, 6:1392–1398, 2021.

\bibitem{Ganose2018}
A.~M. Ganose, S.~Matsumoto, J.~Buckeridge, and D~O. Scanlon.
\newblock Defect engineering of earth-abundant solar absorbers bisi and bisei.
\newblock {\em Chemistry of Materials}, 30:3827--3835, 2018.

\bibitem{Freysoldt2014}
C.~Freysoldt, B.~Grabowski, T.~Hickel, J.~Neugebauer, G.~Kresse, A.~Janotti,
  and C.~G. Van~de Walle.
\newblock First-principles calculations for point defects in solids.
\newblock {\em Rev. Mod. Phys.}, 86:253--305, 2014.

\bibitem{Squires2023}
A.~G. Squires, D.~O. Scanlon, and B.~J. Morgan.
\newblock py-sc-fermi: self-consistent fermi energies and defect concentrations
  from electronic structure calculations.
\newblock {\em JOSS}, 8:4962, 2023.

\bibitem{Lian2021}
W.~Lian, C.~Jiang, Y.~Yin, R.~Tang, G.~Li, L.~Zhang, B.~Che, and T.~Chen.
\newblock Revealing composition and structure dependent deep-level defect in
  antimony trisulfide photovoltaics.
\newblock {\em Nature Communications}, 12:3260, 2021.

\bibitem{Mavlonov2020}
A.~Mavlonov, T.~Razykov, F.~Raziq, J.~Gan, J.~Chantana, Y.~Kawano,
  T.~Nishimura, H.~Wei, A.~Zakutayev, T.~Minemoto, X.~Zu, S.~Li, and L.~Qiao.
\newblock A review of sb2se3 photovoltaic absorber materials and thin-film
  solar cells.
\newblock {\em Solar Energy}, 201:227--246, 2020.

\bibitem{Ma2019}
C.~Ma, H.~Guo, X.~Wang, Z.~Chen, Q.~Cang, X.~Jia, Y.~Li, N.~Yuan, and J.~Ding.
\newblock Fabrication of sb2se3 thin film solar cells by co-sputtering of
  sb2se3 and se targets.
\newblock {\em Solar Energy}, 193:275--282, 2019.

\bibitem{Wen2018}
X.~Wen, C.~Chen, S.~Lu, K.~Li, R.~Kondrotas, Y.~Zhao, W.~Chen, L.~Gao, C.~Wang,
  J.~Zhang, G.~Niu, and J.~Tang.
\newblock Vapor transport deposition of antimony selenide thin film solar cells
  with 7.6% efficiency.
\newblock {\em Nature Communications}, 9:2179, 2018.

\bibitem{Wang2023}
X.~Wang, S.~R. Kavanagh, D.~O. Scanlon, and A.~Walsh.
\newblock Four-electron negative-u vacancy defects in antimony selenide.
\newblock {\em Phys. Rev. B}, 108:134102, 2023.

\bibitem{Kim2020-2}
S.~Kim, J.~A. Márquez, T.~Unold, and A.~Walsh.
\newblock Upper limit to the photovoltaic efficiency of imperfect crystals from
  first principles.
\newblock {\em Energy Environ. Sci.}, 13:1481--1491, 2020.

\bibitem{Kavanagh2022}
S.~R. Kavanagh, D.~O. Scanlon, A.~Walsh, and C.~Freysoldt.
\newblock Impact of metastable defect structures on carrier recombination in
  solar cells.
\newblock {\em Faraday Discuss.}, 239:339--356, 2022.

\bibitem{Nie2021}
R.~Nie, M.~Hu, A.~M. Risqi, Z.~Li, and S.~I. Seok.
\newblock Efficient and stable antimony selenoiodide solar cells.
\newblock {\em Adv. Sci.}, 8:2003172, 2021.

\bibitem{Tiwari2019}
D.~Tiwari, F.~Cardoso-Delgado, D.~Alibhai, M.~Mombrú, and David~J. Fermín.
\newblock Photovoltaic performance of phase-pure orthorhombic bisi thin-films.
\newblock {\em ACS Appl. Energy Mater.}, 2:3878--3885, 2019.

\bibitem{Dahan2013}
N.~Dahan, A.~Jehl, J.~F. Guillemoles, D.~Lincot, N.~Naghavi, and J.-J. Greffet.
\newblock Using radiative transfer equation to model absorption by thin
  cu(in,ga)se2 solar cells with lambertian back reflector.
\newblock {\em Opt. Express}, 21:2563--2580, 2013.

\bibitem{Kim2021}
S.~Kim and A.~Walsh.
\newblock Ab initio calculation of the detailed balance limit to the
  photovoltaic efficiency of single p-n junction kesterite solar cells.
\newblock {\em Applied Physics Letters}, 118:243905, 2021.

\bibitem{Kresse1996}
G.~Kresse and J.~Furthm\"{u}ller.
\newblock Efficient iterative schemes for \textit{ab initio} total-energy
  calculations using a plane-wave basis set.
\newblock {\em Phys. Rev. B}, 54:11169, 1996.

\bibitem{Perdew2008}
J.~P. Perdew, A.~Ruzsinszky, G.~I. Csonka, O.~A. Vydrov, G.~E. Scuseria, L.~A.
  Constantin, X.~Zhou, and K.~Burke.
\newblock Restoring the density-gradient expansion for exchange in solids and
  surfaces.
\newblock {\em Phys. Rev. Lett.}, 100:136406, 2008.

\bibitem{Grimme2010}
S.~Grimme, J.~Antony, S.~Ehrlich, and S.~Krieg.
\newblock A consistent and accurate \textit{ab initio} parametrization of
  density functional dispersion correction (dft-d) for the $94$ elements h-pu.
\newblock {\em J. Chem. Phys.}, 132:154104, 2010.

\bibitem{Schimka2011}
L.~Schimka, J.~Harl, and G.~Kresse.
\newblock Improved hybrid functional for solids: The hsesol functional.
\newblock {\em J. Chem. Phys.}, 134:024116, 2011.

\bibitem{Heyd2003}
J.~Heyd, G.~E. Scuseria, and M.~Ernzerhof.
\newblock Hybrid functionals based on a screened coulomb potential.
\newblock {\em J. Chem. Phys.}, 118:8207--8215, 2003.

\bibitem{Krukau2006}
A.~V. Krukau, O.~A. Vydrov, A.~F. Izmaylov, and G.~E. Scuseria.
\newblock Influence of the exchange screening parameter on the performance of
  screened hybrid functionals.
\newblock {\em J. Chem. Phys.}, 125:224106, 2006.

\bibitem{Pan2018}
J.~Pan, W.~K. Metzger, and S.~Lany.
\newblock Spin-orbit coupling effects on predicting defect properties with
  hybrid functionals: A case study in cdte.
\newblock {\em Phys. Rev. B}, 98:054108, 2018.

\bibitem{Blochl1994}
P.~E. Blöchl.
\newblock Projector augmented-wave method.
\newblock {\em Phys. Rev. B}, 50:17953, 1994.

\bibitem{Mosquera-Lois2022}
I.~Mosquera-Lois, S.~R. Kavanagh, A.~Walsh, and D.~O. Scanlon.
\newblock Shakenbreak: Navigating the defect configurational landscape.
\newblock {\em JOSS}, 7:4817, 2022.

\bibitem{Freysoldt2009}
C.~Freysoldt, J.~Neugebauer, and C.~G. Van~de Walle.
\newblock Fully ab initio finite-size corrections for charged-defect supercell
  calculations.
\newblock {\em Phys. Rev. Lett.}, 102:016402, 2009.

\bibitem{Kumagai2014}
Y.~Kumagai and F.~Oba.
\newblock Electrostatics-based finite-size corrections for first-principles
  point defect calculations.
\newblock {\em Phys. Rev. B}, 89:195205, 2014.

\bibitem{Kavanagh2024}
S.~R. Kavanagh, A.~G. Squires, A.~Nicolson, I.~Mosquera-Lois, A.~M. Ganose,
  B.~Zhu, K.~Brlec, A.~Walsh, and D.~O. Scanlon.
\newblock doped: Python toolkit for robust and repeatable charged defect
  supercell calculations.
\newblock {\em Journal of Open Source Software}, 9:6433, 2024.

\bibitem{Mosquera-Lois2021}
I.~Mosquera-Lois and S.~R. Kavanagh.
\newblock In search of hidden defects.
\newblock {\em Matter}, 4:2602–2605, 2021.

\bibitem{Mosquera-Lois2024}
I.~Mosquera-Lois, S.~R. Kavanagh, A.~M. Ganose, and A.~Walsh.
\newblock Machine-learning structural reconstructions for accelerated point
  defect calculations.
\newblock {\em npj Computational Materials}, 10:121, 2024.

\bibitem{Kim2020}
S.~Kim, S.~N. Hood, P.~van Gerwen, L.~D. Whalley, and A.~Walsh.
\newblock Carriercapture.jl: Anharmonic carrier capture.
\newblock {\em JOSS}, 5:2102, 2020.

\bibitem{database}
C.~López and C.~Cazorla.
\newblock {\em MChX intrinsic point defects. NOMAD}, 2025.

\end{thebibliography}

\end{document}